\documentclass[preprint,showpacs,nofootinbib,prd,aps]{revtex4-1}
\usepackage[utf8]{inputenc}
\usepackage{graphicx,amstext,amssymb}

\begin{document}
\title{Direct photon spectrum and elliptic flow produced from Pb+Pb collisions at $\sqrt{s_{NN}}=2.76$ TeV at the CERN Large Hadron Collider within an integrated hydrokinetic model}

\author{V.~Yu. Naboka$^1$}
%\author{P.~Braun-Munzinger$^{2}$}
\author{Yu.~M.~Sinyukov$^{1}$}
\author{G.M. Zinovjev$^{1}$}
\affiliation{$^1$Bogolyubov Institute for Theoretical Physics,
03680 Kiev,  Ukraine}
%\\
%$^2$ExtreMe Matter Institute EMMI, GSI~Helmholtzzentrum f\"ur~Schwerionenforschung,
%D-64291 Darmstadt, Germany}

\begin{abstract}
The photon transverse momentum spectrum and its anisotropy from Pb+Pb collisions at the CERN Large Hadron Collider energy
$\sqrt {s_{NN}}=2.76$ TeV are investigated within the integrated hydrokinetic model (iHKM). Photon
production is accumulated from the different processes at the various stages of relativistic heavy
ion collisions: from the primary hard photons of very early stage of parton collisions to the thermal
photons from equilibrated quark-gluon and hadron gas stages. Along the way a hadronic medium evolution
is treated in two distinct, in a sense opposite, approaches: chemically equilibrated and chemically
frozen system expansion. Studying the centrality dependence of the results obtained allows us to conclude
that a relatively strong transverse momentum anisotropy of thermal radiation is suppressed by prompt
photon emission which is an isotropic. We find out that this effect is getting stronger as
centrality increases because of the simultaneous increase in the relative contribution of prompt photons
in the soft part of the spectra. The substantial results obtained in iHKM with nonzero viscosity
($\eta/s=0.08$) for photon spectra and $v_2$ coefficients are mostly within the error bars of
experimental data, but there is some systematic underestimation of both observables for the near
central events. We claim that a situation could be significantly improved if an additional photon
radiation that accompanies the presence of a deconfined environment is included. Since a matter of a
space-time layer where hadronization takes place is actively involved in anisotropic transverse flow, both positive contributions to the spectra and $v_2$ are considerable, albeit such an argument needs further research and elaboration.

\end{abstract}

\maketitle

\section{Introduction}

The real photons (and dileptons) measured in relativistic heavy ion collisions were recognized
as unique messengers while probing the new state of produced quark-gluon matter many years ago
\cite{Fein, Shur}. The photons are not strongly interacting and their mean-free path, while traversing
a strongly interacting environment that resulted from hadronic collisions, is large enough to be emitted
almost undistorted after initial collisions. Despite the unqualified acceptance of quantum
chromodynamics (QCD) as the microscopic theory of strong interactions, its application to such complex
reactions as collisions of relativistic heavy ions remains a challenge still far from being resolved
because of the very sophisticated interplay of different momentum scales (hard processes, confinement, etc.)
in the problem. Today it is clear the gauge field theories (and QCD, in particular) being geometrical
in their mathematical nature develop structures with local and global features which can have
a profound impact on the underlying physics (for instance, instantons and monopoles) and the only way of
handling such theories is to construct some effective theories approximating the QCD dynamics in the
particular physical regimes. The integrated hydrokinetic model (iHKM) \cite{iHKM} that we are dealing with
in the present phenomenological analysis clearly benefits from such an approach, because it contains all the stages of matter evolution in A+A collisions, including the early prethermal stage where the thermalization process begins.

This status of underlying theoretical tools for investigating heavy ion collision processes implies that the subsequent development of these studies relies primarily on the search for possible
photon radiation sources \cite{Camp} in collisions and the comparison of their results with the relevant
experimental data accumulated to that point \cite{Shen}. Many important features of such processes
unconnected with QCD, as dictated by another "geometry", i.e., the conditions of the heavy ion
experiments (for example, the form of nuclei, nucleon distribution inside nuclei and its fluctuations,
number of participants in heavy ion collision, centrality of collision, etc.) being unmeasurable
directly in experiments, as a matter of fact are closely associated with some measurable characteristics
of nucleus-nucleus interactions. As to photon (dilepton) production, nowadays it is recognized that
their spectra provide information on the state of the produced system just at the moment of photon
radiation and, hence, can even test some of the QCD calculations. Extensive theoretical studies of
photon production inspired by the unexpectedly large direct-photon yield and their elliptic flow
measured by the PHENIX Collaboration at the BNL Relativistic Heavy Ion Collider (RHIC) in recent years \cite{RHIC_photon} have considerably
increased the possible number of photon radiation sources in order to resolve the "direct-photon flow
puzzle". The latter was observed by the ALICE Collaboration at CERN \cite{ALICE_spectra} as
well.

Indeed, the current experiments keep track of the photons produced by various processes along the
relativistic heavy ion collisions although it is difficult
%sometimes
to disentangle the measured
photon spectra from the particular sources. It concerns the direct photons: prompt photons
initiated by hard parton scatterings and jet fragmentations at the initial stage of interaction, and the
photons from pre-equilibrium hot matter (including radiation from the glasma phase
\cite{Paquet, Ber-Ven, Hees}), from the equilibrated quark-gluon plasma stage, from the jet-tagged conversion and
jet bremsstrahlung \cite{Fries} in the thermal encompassing medium while it is still dense enough, and from the
hadron gas \cite{Rapp} resulting from a quark-gluon system evolution through the cross-over region
\cite{Zahed}. Undoubtedly, a significant contribution to photon production comes from the decay of
hadron resonances after freeze-out \cite{Sinha}. (Fortunately for our analysis in this paper,
the available data of the ALICE experiments have excluded such photons.) Besides, several possible
mechanisms (sources) to increase the photon radiation in heavy ion collisions look quite
realistic theoretically but have not been sufficiently definite in phenomenological predictions.
One of those mechanisms suggests, for example, that a very strong magnetic field created in noncentral heavy
ion collisions can increase the photon radiation owing to the conformal anomaly \cite{Basar}
and synchroton radiation \cite{Zakhar,Tuchin}. Another mechanism contributing significantly to the
observed anisotropy of direct photons refers to a "magnetic bremsstrahlung-like radiation" (synchrotron radiation in modern terminology) of quarks in the collective color field ensuring a confinement
\cite{azimuth-aniz}. Additional productive sources of electromagnetic radiation when the QCD
environment undergoes a confinement have been recently launched in Refs. \cite{Kharz, Camp-1, Pratt,
Itakura}.

This diversity of mechanisms and sources of photon emission in relativistic heavy ion
collisions allows us to hope it is possible to achieve satisfactory description of the direct photon
spectra (yields and anisotropic flows) measured in the ALICE experiments at $\sqrt{s_{NN}} = 2.76$ TeV
\cite{ALICE_spectra,Lohner} by combining these mechanisms in an appropriate way, thereby to advance
significantly in understanding the relevant ways of searching a quark-gluon form of matter and to
provide the proper models with more credibility. By direct photons we mean the production of all
the emission sources excepting contribution from the resonance decays. Actually, they, in turn, can be
subdivided into several types depending on their origin and production time: the prompt photons from the
very initial stage of collision, the photons forming at the thermalization (prethermal) stage, thermal photons from quark-gluon plasma, and the photon radiated
by the expanding hadron matter just after the hadronization of QGP. The integrated hydrokinetic model
\cite{iHKM} is quite efficient in handling with three last sources of direct photons.

The paper is organized as follows. Section \ref{sec:prompt} is devoted to the description of prompt
photon spectra calculation. The brief review of iHKM in its application for modeling the matter
evolution necessary to estimate the thermal photon spectra and $v_2$ is given in Sec.
\ref{sec:iHKM}. Then Sec. \ref{sec:thermal} presents the calculations of thermal photon spectra and
$v_2$. In Sec. \ref{sec:surface} we consider the feasible model of direct-photon radiation that
accompanies the hadronization process (synchrotron radiation) and could contribute significantly to
photon emission. Sec. \ref{sec:total} is devoted to the discussion of the results of the approach
developed. Here the analysis of plotted observables is presented at different parameter values and
for the various scenarios, too. Finally, Sec. \ref{sec:summary} concludes giving an outlook for
possible future developments.

\section{Prompt photon spectra}
\label{sec:prompt}
As mentioned above the prompt photons are emitted at the initial stages of ion collisions, and the
leading order $2\rightarrow 2$ QCD processes (Compton scattering and quark-antiquark annihilation) are
considered as the dominant sources of such photons together with the QCD tagged jets fragmenting into
many final states, some of which include photons. Clearly, their spectra are calculated with the
perturbative QCD (pQCD) although one could see some uncertainty in this point at the energies available at RHIC and LHC
energies where the basic theory (especially for thermal photons) is still strongly coupled. In
further calculations we rely on the results of the experiments \cite{ATLAS_photon, CMS_photon}
demonstrating that the prompt photon spectra scale with the binary nucleon-nucleon collision numbers
$N_{coll}$. It allows us to calculate the spectra for the A+A collisions as a convolution of pQCD
photon spectra in the p+p collision with the number of collisions $N_{coll}$. This number $N_{coll}$
was calculated according to \cite{Ncoll} with the Monte Carlo Glauber code.

The cross section of prompt photon production for the proton-proton collisions can be presented as
\begin{equation}
d\sigma = \sum_{i,j,k}f_i\otimes f_j \otimes d\hat{\sigma}(ij\rightarrow k)\otimes D^{\gamma}_k,
\label{g_s}
\end{equation}
where the summation runs over all possible partonic subprocesses, $f_i$ and $f_j$ are the parton
distribution functions, $D^{\gamma}_k$ is the fragmentation function and $d\hat{\sigma}$ is the
cross section of the corresponding partonic subprocess. The cross-section $d\hat{\sigma}$ is
calculated by the perturbative expansion in the strong coupling constant $\alpha_s$. The total
cross-section depends on the QCD scales $Q_{fact}, Q_{ren},$ and $Q_{frag}$. The factorization scale
$Q_{fact}$ is included in the parton distribution functions $f_i$ and $f_j$, the fragmentation scale
$Q_{frag}$ is included in the fragmentation function $D^{\gamma}_k$ and the renormalization one
$Q_{ren}$ is absorbed by the strong coupling constant. All the scales are set to $Q=0.5~p_T$ in our
calculations because it is well known the smaller proportionality constant leads to better description
of experimental data. We are dealing with the JETPHOX package \cite{JETPHOX} to calculate the
proton-proton spectra for $1$ GeV/c $< p_T < 4$ GeV/c. However, it is hardly consistent with the low-$p_T$ limit of the JETPHOX ability to calculate spectra for the scales $Q=0.5~p_T$. Meanwhile, it was
demonstrated \cite{Paquet} that the change of QCD scale proportionality coefficient leads only to
renormalization of spectra. Thus, in order to calculate low-energy photon spectra for such scales
we first have to calculate the spectra for the scale set to $Q=4.0~p_T$ extending the low-$p_T$ limit, and
then rescale the spectra based on the results for high $p_T$. We use the EPS09 parton distribution
function (PDF) in our calculations (for more details on these PDFs see \cite{EPS09}). We also use the BFG II
fragmentation functions in our calculations \cite{BFG-2} which are represented as tables of values.

\section{Integrated hydrokinetic model}
\label{sec:iHKM}
Estimating the thermal photon spectra and $v_2$ necessarily requires a plausible model of matter
evolution. As such we use the integrated hydrokinetic model \cite{iHKM} throughout this paper, which is
rather efficient for modelling matter evolution from the initial Bjorken time $\tau_0 = 0.1$ fm/$c$ till
the hadronic freezeout. The iHKM depicts such an evolution as consisting of five stages:

1) The initial state is generated by GLISSANDO \cite{gliss,gliss2}. We use a mixed model of wounded
nucleons and binary collisions with a weight coefficient $\alpha$ choosing its value of $\alpha=0.24$
because it successfully describes the experimentally observed correlation between multiplicity and
centrality. The normalization of initial energy density in iHKM is made for centrality $0-5\%$ and
then the normalization coefficients for other centralities are fixed automatically. It also concerns
the $0-20\%$, $20-40\%$ and $40-80\%$ centralities which are treated in this paper. We present the
initial parton/gluon distribution in the form
\begin{equation}
f(t_{\sigma_0},\textbf{r}_{\sigma_0},\textbf{p}) = \epsilon(b;\tau_0, {\bf r}_T)f_0(p).
\label{f0_tot}
\end{equation}
Here the particle distribution in coordinate space $\epsilon(b;\tau_0, {\bf r}_T)$ is set to be
proportional to the relative deposited strength (RDS) calculated by GLISSANDO as described above. The
particle distribution in momentum space is motivated by the Color Glass Condensate (CGC) anisotropic
distribution as
\begin{eqnarray}
f_0(p)=\exp\left(-\sqrt{\frac{(p\cdot U)^2-(p\cdot
V)^2}{\lambda_{\perp}^2}+\frac{(p\cdot
V)^2}{\lambda_{\parallel}^2}}\right),
\label{f0_p1}
\end{eqnarray}
where $U^{\mu}=( \cosh\eta, 0, 0, \sinh\eta)$, $V^{\mu}=(\sinh\eta, 0, 0,\cosh\eta)$. Thus, in the
local rest frame of the fluid element $\eta=0$ and we have
\begin{eqnarray}
f_0(p)=\exp\left(-\sqrt{\frac{p_{\perp}^2}{\lambda_{\perp}^2}+\frac{p_{\parallel}^2}
{\lambda_{\parallel}^2}}\right),
\label{f0_p2}
\end{eqnarray}
where $\lambda_{\perp}$ and $\lambda_{\parallel}$ being in some analogy with the two temperatures, one
in the plane perpendicular to the beam direction and another along the beam direction, correspondingly. We
introduce a parameter $\Lambda=\lambda_{\perp}/\lambda_{\parallel}$ that has meaning as the initial-state momentum anisotropy. We ascertain $\Lambda=100$ in our calculations, just the same as in the
original paper \cite{iHKM} and in accord with the CGC approach.

2) The relaxation model of the prethermal stage \cite{relaxation} describes the continuous transition from
the nonequilibrium state, associated with the time $\tau_0=0.1$ fm/$c$, to the thermalized state,
associated with the time $\tau_{th}=1.0$ fm/$c$. The exact value of $\tau_{th}$ may change, but the
resulting observables are little dependent on this value \footnote{It was shown \cite{tau_th,relaxation}, that during the nonthermal stage, radial flow and elliptic flow in the matter arise even without pressure gradients, just because of system`s finiteness (density gradients) and azimuthal asymmetry. Transformation of the developing collective flows into a Doppler-shifted spectrum and its anisotropy occurs only after thermalization, but the value of the thermalization time is not very important: the system at the pre-thermal stage does not waste time, and collective flows and other effects develop very close to the hydrodynamics rate. In fact, in the model with the prethermal (thermalization) stage, the role of $\tau_{th}$ (used in many earlier studies) is assumed by $\tau_0$ and the value of the latter is critical for observables.} . The relaxation model is motivated by the
proper Boltsmann equations in the integral form. The energy-momentum tensor of the matter at this
stage has a form
\begin{eqnarray}
T^{\mu \nu}(x)=T^{\mu \nu}_{\text{free}}(x){\cal
P}(\tau)+T_{\text{hydro}}^{\mu \nu}(x)[1-{\cal P}(\tau)], \label{tensor}
\end{eqnarray}
where $T^{\mu \nu}_{\text{free}}(x)$ and $T^{\mu \nu}_{\text{hydro}}(x)$ are free and hydrodynamically
evolving components of the energy-momentum tensor. $\cal P(\tau)$ is the probability/weight function
that was chosen in the form \cite{preth, relaxation}
\begin{eqnarray}
{\cal P}(\tau)=  \left (
\frac{\tau_{\text{th}}-\tau}{\tau_{\text{th}}-\tau_0}\right
)^{\frac{\tau_{\text{th}}-\tau_0}
 {\tau_{\text{rel}} }},
 \label{P}
\end{eqnarray}
where $\tau_{rel}$ is the parameter of the model representing the rate of transition from the
nonequilibrium stage to the equilibrated one, and we reckon $\tau_{rel}=0.25$ fm/c in our
calculations \cite{iHKM}. Taking the conservation laws for the total energy-momentum tensor as
$\partial_{;\mu}T^{\mu\nu}_{total}=0$ and assuming that the free-streaming part obeys
$\partial_{;\mu}T^{\mu\nu}_{free}=0$, we obtain the evolution equation as follows
\begin{eqnarray}
\partial_{;\mu}\widetilde{T}^{\mu
\nu}_{\text{hydro}}(x)= - T^{\mu \nu}_{\text{free}}(x)\partial_{;\mu}
{\cal P}(\tau), \label{pre-equation}
\end{eqnarray}
where $\widetilde{T}^{\mu\nu}_{\text{hydro}}(x)=[1-{\cal P}(\tau)]T^{\mu\nu}_{\text{hydro}}(x)$.
The evolution equation for the shear viscosity tensor takes the form \cite{relaxation, iHKM}
\begin{eqnarray}
[1-{\cal P}(\tau)]\left \langle u^\gamma \partial_{;\gamma}
\frac{\widetilde{\pi}^{\mu\nu}}{(1-{\cal P}(\tau))}\right \rangle
=-\frac{\widetilde{\pi}^{\mu\nu}-[1-{\cal
P}(\tau)]\pi_\text{NS}^{\mu\nu}}{\tau_\pi}-\frac {4}{3}
\widetilde{\pi}^{\mu\nu}\partial_{;\gamma}u^\gamma.
\label{pre-viscous}
\end{eqnarray}
Then according to Eq.(\ref{P}) the basic equations for the pre-thermal stage [Eqs. (\ref{pre-equation}) and
(\ref{pre-viscous})] at the (proper) time $\tau=\tau_{th}$ are converted to the equations of
relativistic viscous hydrodynamics in the Israel-Stewart form and the system evolution becomes purely hydrodynamic.

3) The hydrodynamic stage lasts from $\tau_{th}=1.0 fm/c$ to the hypersurface of constant temperature
$T=165$ MeV. This stage is modeled by the same evolution equations as the previous stage but with a
zero source. The equation of state for the matter at this and previous stages develops the form of
the Laine-Schroeder equation of state \cite{Laine}, which provides us with a continuous transition from
a liquid phase to a gaseous one on a hypersurface $T=165$ MeV. The viscosity to entropy ratio is taken
to be equal to its minimal value, $\eta/s=0.08$, and the same for the prethermal stage \cite{iHKM}.

4) At the particlization stage, as noted, the matter evolution is simulated by the hydrodynamic model until the
isotherm with temperature $T_p=165$ MeV. Along this isotherm we utilize a sudden particlization
switching from near locally equilibrated matter evolution to the particle cascade. In this note we
suppose that a hadronization/switching hypersurface is coincident with the hadronization hypersurface
($T_p=T_h=165$ MeV). The hypersurface is built on the computational grid using the Cornelius routine
\cite{convert}. The well-known Cooper-Frye formula
\begin{eqnarray}
p^0 \frac{d^3 N_i(x)}{d^3 p}=d\sigma_{\mu}p^{\mu}f(p\cdot u(x),T_h,\mu^{(i)}_h)
\label{Cooper-Frye1}
\end{eqnarray}
is used to convert the fluid to the particle cascade. Grad's 14-momentum anzats is used to account
the viscous corrections to the particle distribution function.

5) Next is the final hadronic stage. In the original iHKM, the further evolution of the hadronic cascade is
simulated by ultrarelativistic quantum molecular dynamics (UrQMD) \cite{urqmd}. However, a calculation
of photon radiation at this stage accounting for the hadron reactions in UrQMD is currently not
possible and we consider instead the two variants of the hydrokinetic evolution below $T_h=165$ MeV.
The first one suggests continuation of hydrodynamic evolution of the hadron matter as
the {\it chemically equilibrated} expansion. As a result, the field of collective velocities and
temperatures for hadron matter will be defined explicitly, and then one can estimate photon radiation
from the expanding hadron medium. For this purpose one can use the known results for the hadron
reactions with the photon radiation in a resting thermal hadron system. A disadvantage of such an
approach is that at relatively small temperatures the chemical equilibrium is certainly violated. The
second variant is to utilize the original version of hydrokinetic model HKM \cite{original HKM} with
continuous particlization, violation of the local thermal and chemical equilibration and {\it chemically
frozen} (all inelastic reactions except for the resonance decays are forbidden) evolution. It again
allows one to estimate the 4-velocities, temperatures and chemical potentials (for each hadron!) in an
expanding hadron medium. We are dealing with both opposite variants of the hadronic matter evolutions
to estimate the upper and lower limits of hadron spectra and $v_2$ that such rough approximations
bring.

\section{Prethermal and thermal photon spectra}
\label{sec:thermal}
As it was mentioned in previous section, the role of prethermal stage is mostly in formation of initial conditions~--- the energy density profile and field of initial velocities (both longitudinal and anisotropic transverse ones)~--- for viscous hydrodynamic expansion of quark-gluon plasma. The relative contribution of the photons at comparatively low
transverse momenta $1 < p_T< 4$ GeV/c from this stage of the matter evolution, $\tau = 0.1$~--  $1$ fm/c, is relatively small, a couple of percents of all thermal photons. Since in the relaxation model this stage is represented by two components: parton free streaming, which dies out with time, and the hydrodynamic one, that is gradually forming, we consider approximately the photon radiation from this stage as coming only from a hydrodynamic component with the corresponding wait, $1-{\cal P}(\tau)$, and equation of state and viscosity such as in QGP forming .

Now analyzing the particular LHC data for direct-photon spectra and $v_2$ at
transverse momenta $1 < p_T< 4$ GeV/c we recognize that the dominant contribution to such spectra
comes from the thermal radiation of the hot QGP phase and subsequent hadron gas. The pQCD is a main tool
of all data analyses while calculating the photon emission due to the leading order $2\rightarrow 2$
processes from the hot QGP \cite{Arnold}. However, it was shown \cite{Gelis} that the considerable
contribution to photon emission comes also from the higher-order collinear processes and,
moreover, it is parametrically of the same order as the $2\rightarrow 2$ processes. This mechanism is
quite similar to that for the photon radiation from hard quarks \cite{Zakhar} and to the induced
gluon radiation from fast partons \cite{Zakhar2}. The data analyses based on the standard formulas of
\cite{Arnold} calculated for the fixed QCD coupling constant awaken a lively interest in such
calculations with the running coupling constant (and changing the thermal quark mass). It looks like
a quite relevant task, at least for the energy range covered by RHIC and LHC where the QGP, as we know,
is strongly coupled.

For the further calculations here to make more realistic estimates of the matter evolution we address
again the iHKM. The argument of great importance to do so is the serious statement that iHKM is well
suited to describing many other observables such as the hadron yields, particle spectra, anisotropic flows, and pion and kaon interferometry radii at different centralities. Moreover, in our
present calculations we are handling iHKM with the same set of parameters as used for
describing almost all bulk hadron observables in the past (see \cite{iHKM}). Thus, in estimating
the thermal emission from the QGP phase we draw attention to the expression of \cite{Arnold} with the number of quarks
$n_f=3$ as for the LHC data
\begin{equation}
\frac{d^7N}{d^3\textbf{k}d^4\textbf{x}}=\frac{\nu_e(\left|\textbf{k}\right|)}{(2\pi)^3}
\label{spectra1}
\end{equation}
It was derived assuming an equilibrated quark-gluon plasma with a four-velocity $\textbf{u}=(1,0,0,0)$
in the vicinity of each space-time point. Here $\nu_e(|\textbf{k}|)$ denotes the spontaneous emission rate
for photons of a given momentum $\textbf{k}$. Its relativistically invariant form that we utilize
looks like the following
\begin{equation}
k^0\frac{d^7N}{d^3\textbf{k}d^4\textbf{x}}={k}\cdot {u}~\frac{\nu_e(k\cdot {u})}{(2\pi)^3},
\label{spectra2}
\end{equation}
and for the (photon) midrapidity $|y|<0.5$ we have
\begin{equation}
\frac{dN}{2\pi k_Tdk_T}=\sum_i\frac{1}{(2\pi)^4} \Delta^4 V(x_i)\int_0^{2\pi} {k}\cdot {u(x_i})\nu_e({k}\cdot {u(x_i}))d\phi,
\label{spectra_qgp}
\end{equation}
where the summation is made over all cells of the 3-dimensional computational grid (these dimensions
are $\tau,x,y$, as far as we use the boost-invariant model),
$\Delta^4 V=\tau \Delta \tau\Delta x\Delta y\Delta\eta$ is a volume of the single cell, and $\eta$
is the space-time rapidity.

As Eq. (\ref{spectra_qgp}) is used to describe the emission from the QGP we apply it to the grid
cells with a temperature $T>T_{h}=165$ MeV only. For the cells with the temperature lower than $T_{h}$
we use the same formula (\ref{spectra_qgp}), but with $\nu_e(\textbf{k})$ corresponding to the hadron
emission. The latter includes many channels which are successfully absorbed in the calculations. For
the photon emission from a meson gas we follow \cite{Turbide}, and for the emission with a specific
behaviour of the $\rho$-meson self-energy we use the parametrization from \cite{Heffernan}. The
photons originated by the reactions $\pi+\rho\rightarrow\omega+\gamma,
\rho+\omega\rightarrow\pi+\gamma, \pi+\omega\rightarrow\rho+\gamma$ are taken from \cite{Rapp}
(including the t-channel emission of $\omega$). And finally we include $\pi\pi$ bremmstrahlung
considered in \cite{Heffernan}. As we mentioned above, since there is no possibility of calculating photon emission in UrQMD, we use the background hydrodynamics of hadron fluid that is utilized in the original HKM \cite{original HKM, HKM}. In the original HKM model there is \emph{no} sudden thermal freeze-out, like in UrQMD:  after hadronization at 165 MeV,  particles leave expanding hadron fluid continuously, and hadron spectra are close to those in the picture with the hadron cascade at the latest stage of the evolution \cite{HKM}. So, to see uncertainties of such an approximation in the temperature range
$100$ MeV $< T < 165$ MeV, we use the two approaches as mentioned in Sec.\ref{sec:iHKM}. One is based
on chemically equilibrated hydrodynamic evolution past the hypersurface temperature
$T = 165$ MeV down to the isotherm $T = 100$ MeV, and the second one is used as a background of continuous emission and is a chemically frozen (only resonance decay allowed) hydrodynamic expansion \cite{original HKM, HKM}.

\section{ Photon radiation at hadronization stage}
\label{sec:surface}

It was suggested \cite{azimuth-aniz} that a considerable additional contribution to the photon production
could come due to the boundary bremsstrahlung resulting from the interaction of escaping quarks with the
collective confining color field at the surface of the QGP. Obviously, such a mechanism of "magnetic
bremsstrahlung-like radiation" (synchrotron radiation in modern terminology) should manifest itself
noticeably in the observed anisotropy of direct photons. We develop this idea supposing that a
specific photon radiation takes place during the process of hadronization. Note, the different mechanisms of additional photon radiation from a confining process are also proposed in Refs.\cite{Kharz, Camp-1,
Pratt, Itakura}.) Trying to give more credibility to such mechanisms we speculate here adapting the
phenomenological prescription for describing the photon emission from the hadronization space-time layer in the cross-over scenario at the LHC.

Let us describe an additional ``hadronization'' photon production by some emission function $G_{hadr}$
\begin{equation}
\frac{d^3N_{\gamma}}{d^3p}=\int dtd^3r~G_{hadr}(t,{\bf r},p)
\label{G_hadr}
\end{equation}
Let us find at each  $({\bf r},p)$ the temporal point $t({\bf r}, p)$ of the maximal emission of the photons with 4-momentum $p$. These points form the hypersurface $\sigma$: $t_{\sigma}({\bf r}, p)$.
Let us pass to new variables which include this saddle point \cite{var-1,var-2}, namely, ${\bf x}={\bf r}+\frac{{\bf p}}{p^0}t_{\sigma}({\bf r},p)$. Then, using the corresponding Jacobian, and presenting
the photon emission function in the saddle point approximation, $G_{hadr}\approx F(t,{\bf x},p)\exp(-\frac{(t-t_{\sigma})^2}{2D^2})$, with some function $F$, which has smooth dependence on $t$, one can write:
\begin{equation}
\frac{d^3N_{\gamma}}{d^3p}=\int d^3x \left|1-\frac{{\bf p }}{p^0}\frac{\partial t_{\sigma}}{\partial {\bf x}}\right|\int dt F(t_{\sigma}, {\bf x}, p) \exp\left(-\frac{(t-t_{\sigma})^2}{2D_c^2(t_{\sigma}, {\bf x}, p)}\right)
\label{add}
\end{equation}
Here $t_{\sigma}= t_{\sigma}({\bf x},p)$. If one assumes that the hypersurface of maximal photon emission due to the hadronization process corresponds to the isotherm $T_h$ (in our model $T_h=165$ MeV) for all momenta $p$, then Eq. (\ref{add}) can be written in the invariant form
\begin{equation}
p^0\frac{d^3N_{\gamma}}{d^3p}=\int_{\sigma_h}d^3\sigma_{\mu}(x)p^{\mu}F\left(p\cdot u(x),T_h\right)D_c\left(p\cdot u(x),T_h\right)\theta(d\sigma_{\mu}(x)p^{\mu}).
\label{cooper_frye}
\end{equation}
Here chemical potentials are made to be equal to zero at the hadronization temperature at the LHC energies and $\theta(z)$ is the Heaviside step function that is designed to exclude a negative contribution to
the spectra from possible not-space-like parts of the hadronization hypersurface. While the function $F$ is defined by the basic confinement and hadronization properties in the medium related to unity of volume and unity of time, the temporal width $D_c$ depends on centrality $c$ since the duration of hadronization process depends on rate of expansion, which is higher in noncentral collisions (initial transverse gradients are larger in narrower system) thereby reducing $D_c$ for more peripheral collisions.

		Our main goal here is to include an additional radiation
mechanism in the simplest phenomenological version in order to test to what extent this mechanism can improve
 the description of the photon spectra and $v_2$ coefficients. The simplest treatment also allows one to
eliminate some theoretical uncertainties of different approaches concerning the hadronization process which are usually hidden in tuning parameters. So we suppose in Eq. (\ref{cooper_frye}) that the $FD$ function has the thermal-like form $f_{\gamma}^{eq}$ function
\begin{equation}
F D_c = d_c\gamma_{hadr}~f_{\gamma}^{eq}\left(p\cdot u(x),T_h\right)= \gamma_{hadr}d_c\frac{1}{(2\pi)^3}\frac{g}{\exp\left(p\cdot u(x)/T_h\right)-1},
\label{surface}
\end{equation}
where $p\cdot u \equiv p^{\mu}u_{\mu}$, $g=2$ and $T_h = 165$ MeV is the effective hadronization
temperature, and $u(x)$ is the collective 4-velocity. The value of $\gamma_{hadr}$ is defined by the basic hadronization process, while $d_c \propto \left\langle D\right\rangle$  is defined by the temporal width of the hadronization process that depends on the collision centrality $c$ and is reduced with an increase of impact parameter. We take the value of $\alpha\equiv d_c\gamma_{hadr}=0.02$ for most central events as providing the better description of total photon spectra and momentum anisotropy. If we use the same value 0.02 for $\alpha$ also for non-central interval in $c$, then the results become worse compared to the calculations without this additional contribution  (although they are still within the error bars). It confirms our remark above, that the temporal width of hadronization is smaller in noncentral collisions.

\section{Results and discussion}
\label{sec:total}
We calculate the total photon spectra as a sum of thermal photon spectra, prompt photon spectra,
(which may not be neglected as seen from the results), and the spectra associated with
hadronization processes. The latter summand is included for the most central collisions only. Besides,
we use the same set of parameters that is considered in \cite{iHKM} to be optimal, i.e. minimal
viscosity, $\eta/s=0.08$, high initial-state momentum anisotropy, $\Lambda=100$, relatively
small relaxation time, $\tau_{rel}=0.25$ fm/c, compared to the time of thermalization, $\tau_{th}=1$
fm/$c$, and early initial-state formation time, $\tau_0=0.1$ fm/c. It is also worth noting that as far
as iHKM includes the prethermal stage (just after the initial collision of nucleons or partons in CGC
approach and before the start of the hydrodynamic stage) the resulting photon spectra includes the emission from the
prethermal stage as well.

\begin{figure}
     \centering
     \includegraphics[width=1.0\textwidth]{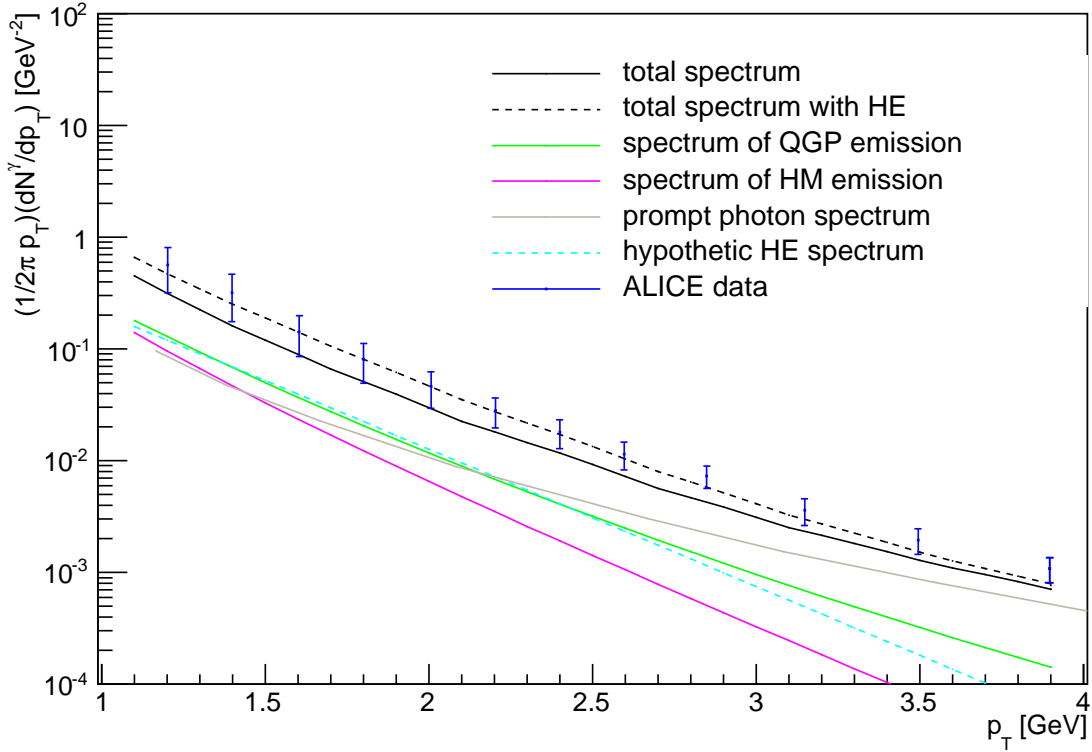}
     \caption{\small
Total direct-photon spectra in iHKM: thermal QGP + thermal HM  + prompt + hadronization emission (HE). Centrality is 0-40\%. Experimental results are taken from \cite{ALICE_spectra}.}
\label{fig:total040}
\end{figure}

\begin{figure}
     \centering
     \includegraphics[width=1.0\textwidth]{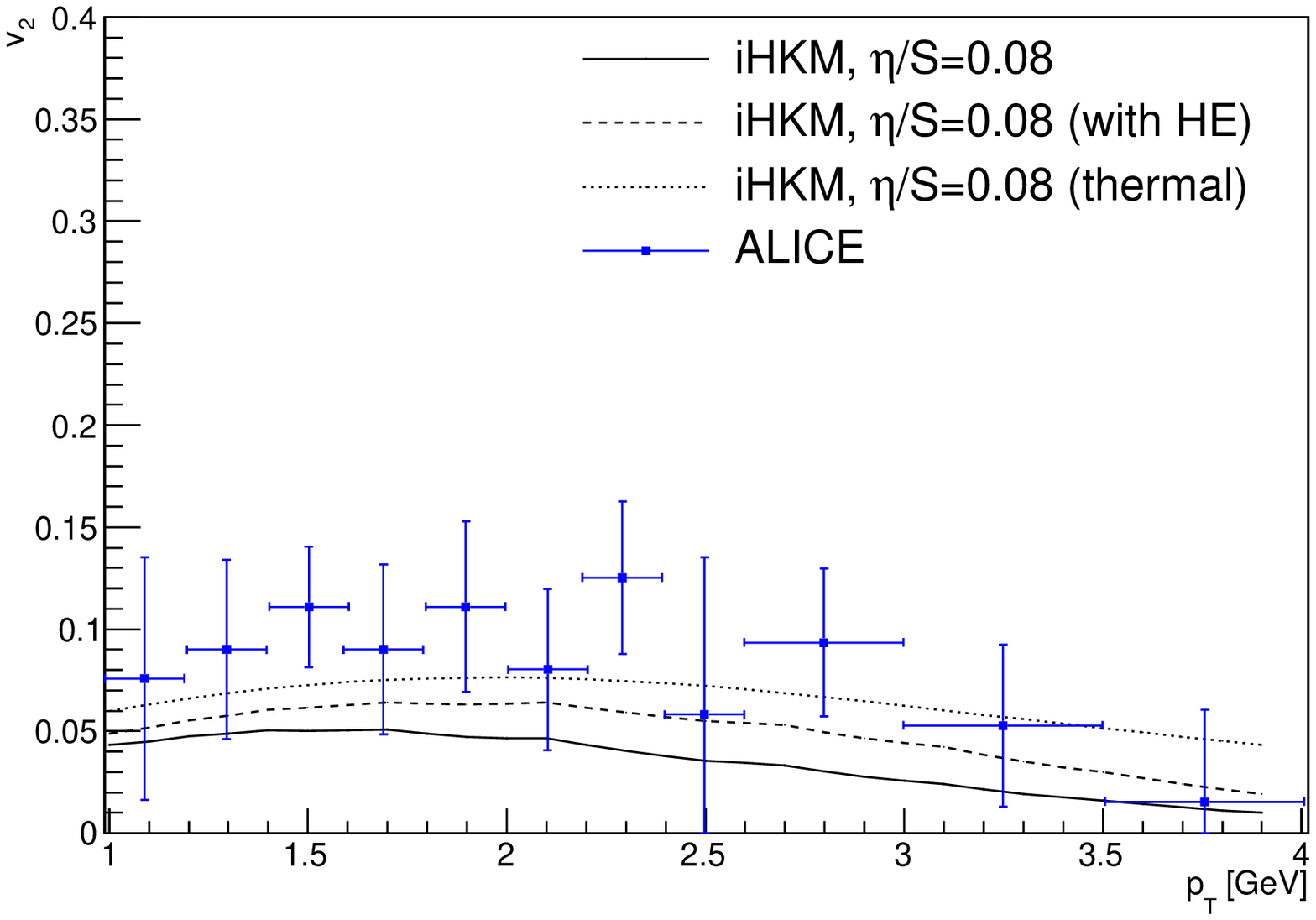}
     \caption{\small
Photon momentum anisotropy $v_2$-coefficient for 0-40\% centrality. The results including the
hadronization emission (HE) and results for thermal photons only (without HE)
are also presented. Experimental results are taken from \cite{Lohner}.}
\label{fig:v2_040}
\end{figure}

We start with the scenario when an expansion of hadron matter after particlization or hadronization is
chemically equilibrated. The resulting total spectra for 0-40\% centrality is shown in
Fig. \ref{fig:total040} together with all its components and experimental results. In this approach
a hydrodynamic stage lasts from $T_h=165$ MeV until $T=100$ MeV to describe the photon
emission from the expanding hadron matter (HM). We are doing it in the two alternative ways explained
in Sec. \ref{sec:iHKM}, for chemically equilibrated and chemically frozen expansion of the system.
Figure \ref{fig:total040} demonstrates that the prompt photon spectrum gives less impact to the total spectrum than the
thermal one, which is formed by the QGP+HM system for small photon transverse momenta, while the prompt photon spectrum starts to
dominate for $p_T > 3-4$ GeV/c. The additional  photon emission (HE) due to thermalization process from the temporary
narrow hadronization space-time layer (approximated by the hypersurface of constant temperature $T_h =165$ MeV), gives
a quite noticeable contribution, as one can see from Fig. \ref{fig:total040}. The intensity here is
defined by the weight value $\alpha = 0.02$ in Eq. (\ref{surface}) that is chosen to provide the best fit
of the total photon spectra at the most central interval. The momentum dependence of the photon anisotropic flow coefficient $v_2$
under the same conditions is plotted in
Fig.\ref{fig:v2_040}. It is curious that the additional emission corrects the total
photon spectra well and improves the description of the $v_2$-coefficients. It suggests that the matter
concentrated in the hadronization space-time layer ($\approx$ hadronization hypersurface) is actively
involved in strong anisotropic flow.

\begin{figure}
     \centering
     \includegraphics[width=1.0\textwidth]{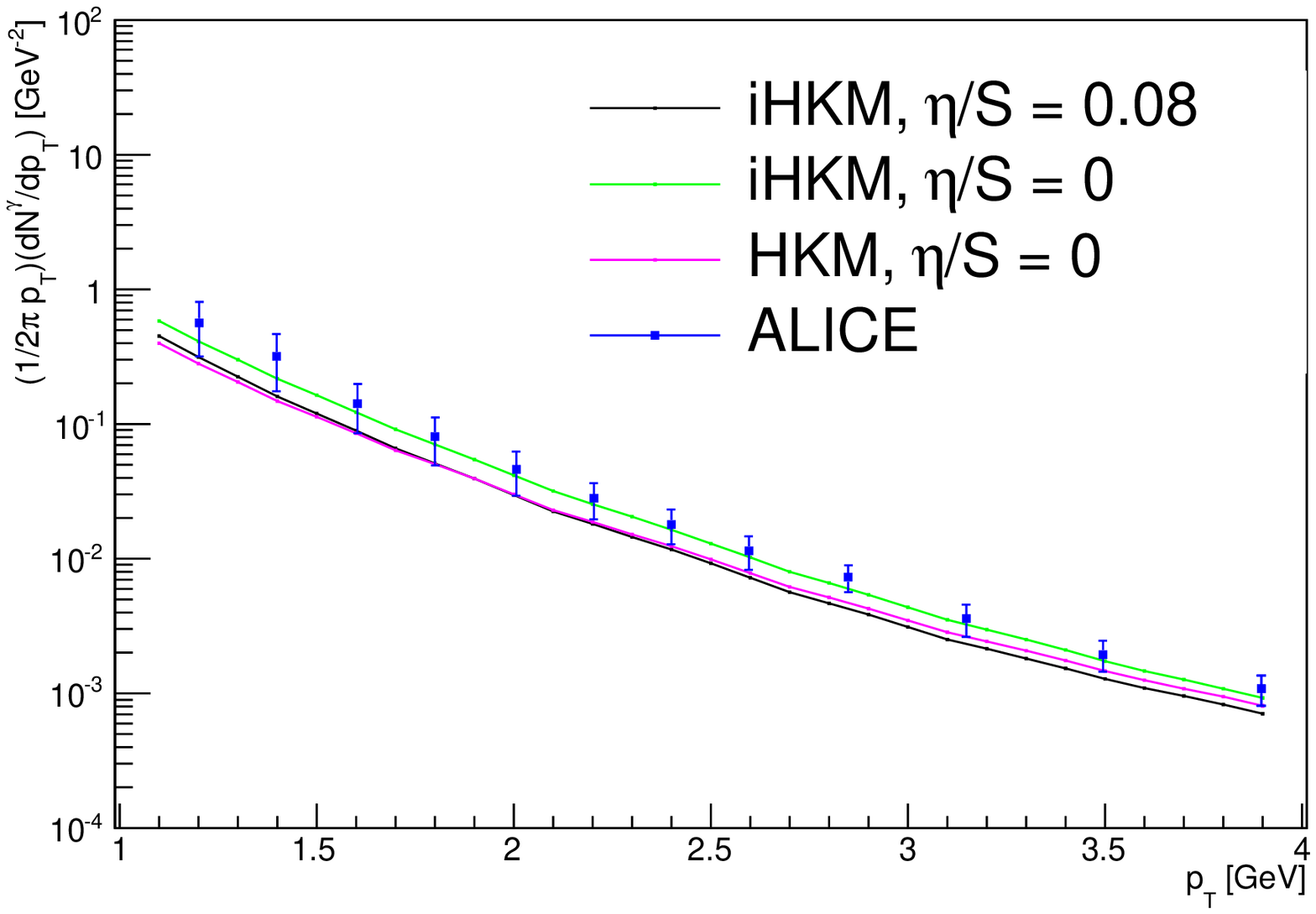}
     \caption{\small
Total photon spectra for 0-40\% centrality for the different models: iHKM chemically equilibrated
viscous ($\eta/s = 0.08$), ideal ($\eta/s = 0$) and HKM chemically frozen at the hadron stage with
continuous transition from hydrodynamics to hadron gas. HE contribution is not included. Experimental
results are taken from \cite{Lohner}.}
\label{fig:total040_comp}
\end{figure}

\begin{figure}
     \centering
     \includegraphics[width=1.0\textwidth]{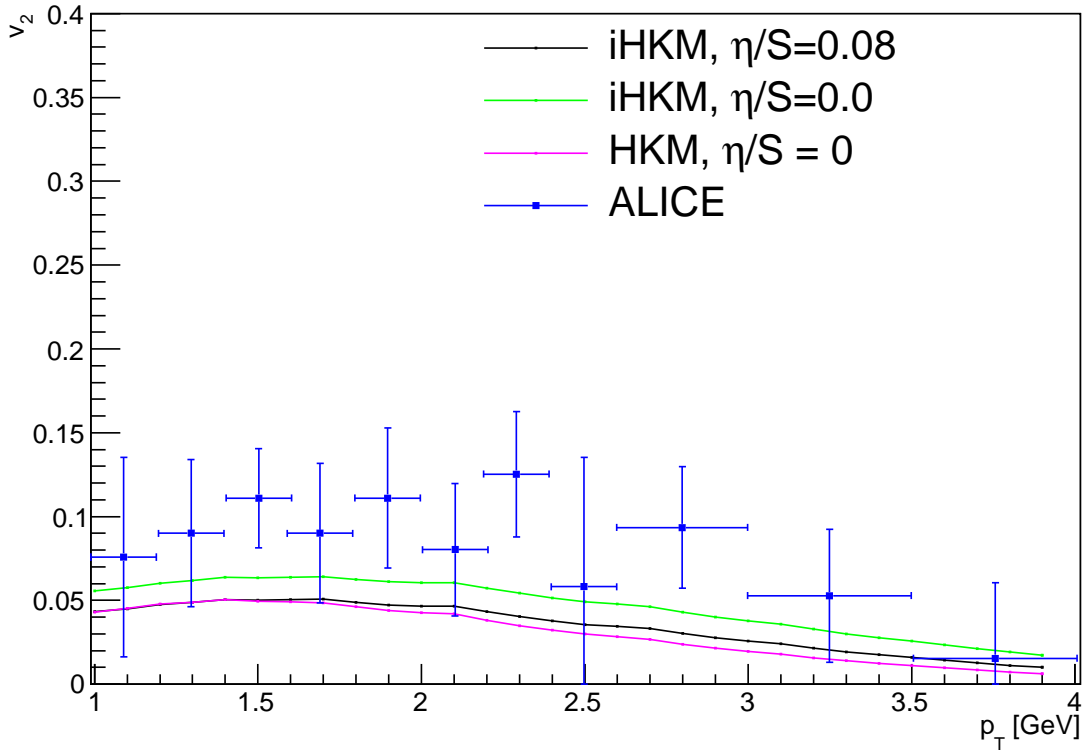}
     \caption{\small
Photon momentum anisotropy, $v_2$, for 0-40\% centrality for different models: iHKM chemically
equilibrated viscous ($\eta/s = 0.08$), ideal ($\eta/s = 0$) and HKM chemically frozen at the hadron
stage  with continuous transition from hydrodynamics to hadron gas. HE contribution is not included.  Experimental results are taken from \cite{Lohner}.}
\label{fig:v2_040_comp}
\end{figure}

Now in what follows we ignore the synchrotron photon emission and analyze the dependence of the results
on the other factors. It is worthwhile to notice that changing the model parameters leads to the
renormalization of initial energy density. This renormalizations turns out to be mandatory for properly describing
all-charged-hadron multiplicity vs centrality (see details in \cite{iHKM}). The proper factor plays an
important role in reaching the best results for the total photon spectra in
Fig. \ref{fig:total040_comp} when the viscosity effect is ignored, $\eta/s=0$. Due to the
renormalization, initial energy density for perfect hydro-evolution is considerably larger than its
value for the viscous model (see table 1 in \cite{iHKM}), and as the result of longer QGP+HM
expansion, the thermal photon spectra becomes larger. One can see also from
Fig. \ref{fig:total040_comp} that the change of hadron matter evolution scenario from chemically
equilibrated to chemically frozen \cite{original HKM, HKM} expansion does not influence, practically,
the result. The similar conclusion one can get from Fig. \ref{fig:v2_040_comp} for $v_2$
coefficients.

\begin{figure}
     \centering
     \includegraphics[width=1.0 \textwidth]{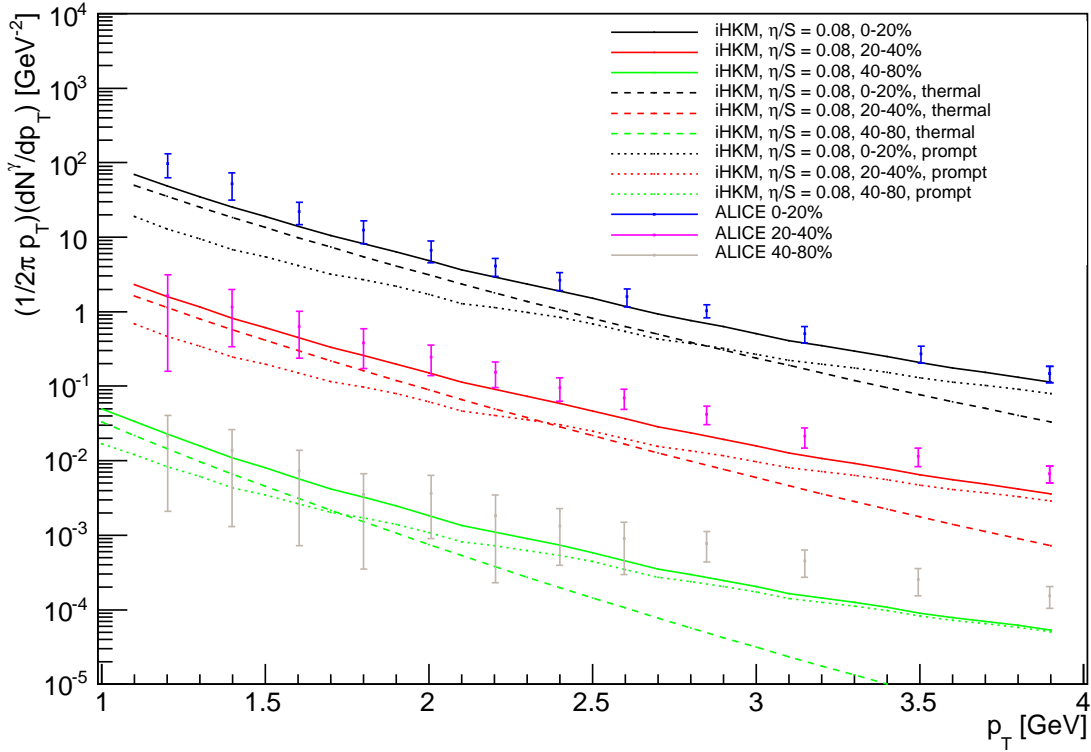}
     \caption{\small
Total direct-photon spectra for different centralities. Viscous ($\eta/s = 0.08$) iHKM model is used
in these calculations. The spectra for 0-20\% centrality is multiplied by a factor of 100 and
spectra for 20-40\% centrality is multiplied by a factor of 10. HE contribution  is not included. QGP+hadron (thermal)
and prompt photon contributions are shown separately. Experimental
results are taken from \cite{ALICE_spectra}.}
\label{fig:spectra_centrality}
\end{figure}

\begin{figure}
     \centering
     \includegraphics[width=1.0\textwidth]{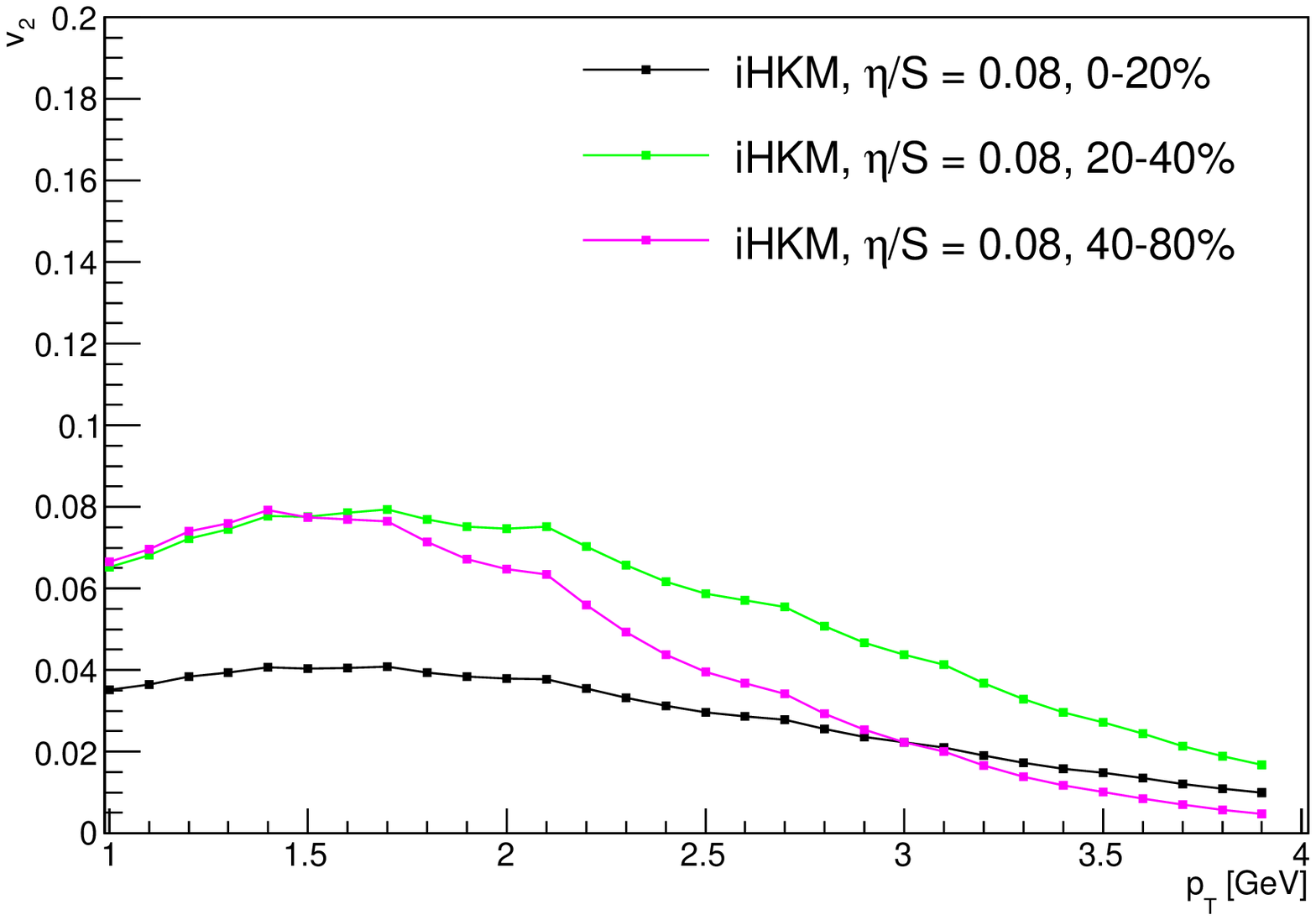}
     \caption{\small
Anisotropic flow coefficient $v_2$ for all direct photons for different centralities. Viscous ($\eta/s = 0.08$) iHKM model is used in these calculations. HE contribution is not included.}
\label{fig:v2_centrality}
\end{figure}

We also investigate the centrality dependence of the results. The corresponding spectra for 0-20\%,
20-40\% and 40-80\% are shown in Fig. \ref{fig:spectra_centrality} in comparison with ALICE data. The
results for $v_2$ are plotted in Fig. \ref{fig:v2_centrality}. One can see how strongly the transverse
momentum anisotropy of the thermal QGP + HM emission is suppressed by the prompt photon emission
(which is momentum isotropic) at different centralities. The more peripheral the collisions, the larger is
relative contribution of prompt photons in the soft part of the spectra. We address these predictions for
upcoming experimental data. Note that the direct-photon data from Au+Au collisions at RHIC by PHENIX are available for three centrality bins. Thus, we are planning to calculate the direct-photon spectra and $v_2$ at RHIC and compare their results with the experimental data for different centrality bins and make a firm conclusion from there. Such a work is in preparation now.

\section{Summary}
\label{sec:summary}

In this paper the photon emission and its transverse momentum anisotropy are investigated for heavy
ion collisions at the LHC energy $\sqrt{s_{NN}} = 2.76$ TeV within the integrated hydrokinetic model.
This model is used with the same parameter set that describes reasonably well almost all bulk
hadronic observables, and includes practically all the stages of collision such as an initial
nonequilibrium state formation, a prethermal stage, a hydrodynamic quark-gluon stage, and subsequent hadron
matter evolution. The prompt photons coming from very initial stage of A+A collisions are taken into
consideration as well. The hadron medium evolution is treated in this work in the two different
(opposite) approaches: a chemically equilibrated and a chemically frozen system expansion.
Both scenarios lead to quite similar results.

The main results obtained within iHKM with nonzero viscosity, $\eta/s=0.08$, for photon spectra and
$v_2$-coefficients do not contradict the experimental data available. They are mostly within the
error bars, but there is some systematic underestimation for both observables for the near central
events. However, the situation is getting better in two ways. The first one corresponds to perfect
$\eta/s=0$ hydrodynamic evolution. It seems to be nonrealistic because
$\eta/s=\frac{1}{4\pi}\approx 0.08$ is the minimal possible ratio of shear viscosity to entropy
density. Another way to get a relatively well description of the photon momentum spectrum and its
anisotropy is to suppose that there is an additional photon radiation accompanying the hadronization
processes. Since the matter in space-time layer, where hadronization occurs, is actively involved in
anisotropic transverse flow, the contributions to the spectra and $v_2$ are significant. This source
of photon radiation does not look too exotic and needs to be investigated theoretically and
especially phenomenologically in the context of kinetic description of heavy ion collisions.
A study of the centrality dependence of the results demonstrate that relatively strong  transverse
momentum anisotropy of the thermal QGP + HM emission is suppressed by the prompt photon emission
that is momentum isotropic. We have found that this effect strengthens with increasing centrality because the relative contribution of prompt photons in the soft part of the spectra grows with
centrality. We address these predictions for upcoming experimental analysis.

\begin{acknowledgments}
The research was carried out within the scope of the EUREA: European Ultra Relativistic Energies
Agreement (European Research Network "Heavy ions at ultrarelativistic energies"), Agreement F-2018 of National Academy of Sciences (NAS) of Ukraine.
The work is partially supported by the NAS of Ukraine Targeted research program "Fundamental research on high-energy physics and nuclear physics (international cooperation)".
\end{acknowledgments}

\end{document}